\documentclass[prl,amsfonts,floatfix,superscriptaddress,twocolumn,longbibliography]{revtex4-1}
\usepackage{mathbbol}
\usepackage{graphicx,color}
\usepackage{amsmath,amssymb,bm,amsthm,amstext}
\usepackage{mathtools}
\usepackage{simplewick}
\usepackage{braket}

\usepackage{xfrac}
\usepackage{tikz}
\usepackage{circuitikz}
\usepackage{epstopdf} 
\usepackage{times}
\usepackage{dsfont}
\usepackage[normalem]{ulem}

\usepackage{bm}
\usepackage{mathrsfs}

\newcommand{\Tr}{\operatorname{Tr}}

\renewcommand{\H}{\mathcal{H}}

\newcommand{\totimes}{\vcenter{\hbox{$\scriptstyle\otimes$}}}
\newcommand{\sigmaz}{\sigma_{\text{\tiny Z}}}
\newcommand{\sigmax}{\sigma_{\text{\tiny X}}}
\newcommand{\sigmay}{\sigma_{\text{\tiny Y}}}
\newcommand{\e}{\text{e}}

\newcommand{\Cz}{\mathsf{C}^\mathsf{z}}
\newcommand{\Cx}{\mathsf{C}^\mathsf{x}}

\begin{document}
\title{Experimental Detection of the Correlation R\'enyi Entropy in the Central Spin Model}

\author{Mohamad Niknam}
\address{Institute for Quantum Computing, University of Waterloo, Waterloo, ON, Canada, N2L3G1}
\address{Department of Physics, University of Waterloo, Waterloo, ON, Canada, N2L3G1}
\author{Lea F. Santos}
\address{Department of Physics, Yeshiva University, New York City, New York, 10016, USA}
\author{David G. Cory}
\address{Institute for Quantum Computing, University of Waterloo, Waterloo, ON, Canada, N2L3G1}
\address{Department of Chemistry, University of Waterloo, Waterloo, ON, Canada, N2L3G1}

\date{\today}

\begin{abstract}
We propose and experimentally measure an entropy that quantifies the volume of correlations among qubits. The experiment is carried out on a nearly isolated quantum system composed of a central spin coupled and initially uncorrelated with 15 other spins. Due to the spin-spin interactions, information flows from the central spin to the surrounding ones forming clusters of multi-spin correlations that grow in time. We design a nuclear magnetic resonance experiment that directly measures the amplitudes of the multi-spin correlations and use them to compute the evolution of what we call correlation R\'enyi  entropy. This entropy keeps growing even after the equilibration of the entanglement entropy. We also analyze how the saturation point and the timescale for the equilibration of the correlation R\'enyi  entropy depend on the system size.
\end{abstract}

\maketitle

Which microscopic entropy can capture the changes undergone by an isolated quantum system that evolves in time? The von Neumann entropy for the entire density matrix of the system is not an appropriate choice, because it is constant in isolated systems.  A common approach is to trace out part of the system and resort to the entanglement entropy, which quantifies the degree of entanglement between the traced-out part and the remaining subsystem. Despite the challenges presented by this quantity, it has been experimentally measured in a system with 3 superconducting qubits after tomographically reconstructing the evolved density matrix~\cite{Neill2016}, in a Bose-Hubbard system with 6 cold atoms and site-resolved number statistics~\cite{Kaufman2016}, and in a chain with 20 trapped ions where the entropy of subsystems with up to 10 ions is obtained through randomized measurements~\cite{Brydges2019}. The entanglement entropy is bounded by the quantum Fisher information. This quantity offers a way to detect the flow of information~\cite{Lu2010,Garttner2018} and has been experimentally measured with trapped ions~\cite{Smith2015} and in nuclear magnetic resonance (NMR)~\cite{Niknam2020}. 

Another entropy that has received more theoretical than experimental attention is the participation R\'enyi entropy, which measures the spread in time of a non-stationary state in the Hilbert space. The system is usually prepared in a certain basis vector and the entropy is computed by summing the squares (or larger powers) of the probabilities for finding the system in its initial quantum state and in each one of the other basis vectors~\cite{Flambaum2001b}. For both the participation R\'enyi entropy and the entanglement entropy, common questions include the conditions for linear or logarithmic growths in association with quantum chaos~\cite{Asplund2014,Bianchi2018,Santos2012PRL,Vidmar2017b,Borgonovi2019R} or the transition to many-body localization~\cite{Znidaric2008,Bardarson2012,Torres2017,Lukin2019,DaumannARXIV}, comparisons between their saturation values and thermodynamic entropies in studies of thermalization~\cite{Santos2012PRER,Kaufman2016}, and analytic predictions for the spread of entanglement~\cite{Page1993,Vidmar2017a,Alba2017,CalabreseARXIV}.  One of the differences between the two entropies is that the participation R\'enyi entropy is extensive in the Hilbert space size of the composite system, while the maximum value of the entanglement entropy is not if the size of the subsystem does not change.

In this work, we propose a third alternative that we measure employing NMR techniques. NMR has been used to investigate questions in many-body quantum dynamics, such as many-body localization~\cite{Wei2018}, prethermalization~\cite{Wei2019,Yin2020}, and the scrambling of quantum information~\cite{Niknam2020,Li2017,Sanchez2020}. Our experiment demonstrates that NMR platforms are also testbeds for the analysis of entropy growth.

Our entropy quantifies the growth of the volume of correlations as information flows from a central spin (qubit) to its surrounding spins. As devices with ever larger numbers of qubits become operational, a detailed picture of how quantum information flows and how the dynamics saturates are essential for designing and controlling quantum processors. This understanding is also necessary for classical simulations, which become impracticable under a substantial growth of correlations.

 \begin{figure}[htb]
 	\centering
 	\includegraphics*[scale=0.45]{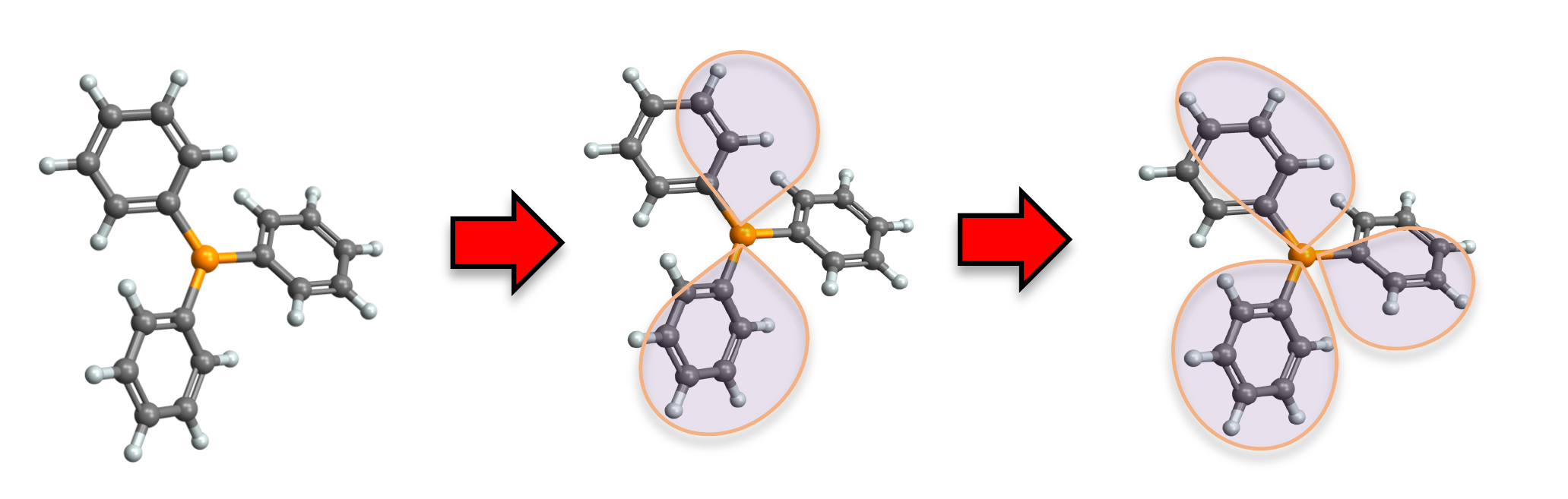}	   	 	 	 	 	 	
 	\caption{Schematic illustration of the flow of information initially contained in the central spin (orange circle) to the surrounding 15 spins. Each shaded area indicates a cluster of correlated spins.
 	} 	
\label{fig01}
 \end{figure}
 
In our sample, the central spin is initially polarized and coupled with 15 unpolarized surrounding spins. This composite system is at room temperature and nearly isolated from external environments. The experiment employs two main ingredients available to solid-state NMR. One is the possibility to coherently average out the interactions among the surrounding spins, so that the remaining effective Hamiltonian contains only the interactions between those spins and the central one. Due to these couplings, as we sketch in Fig.~\ref{fig01}, information that is initially concentrated in the central spin (orange circle) flows to the surroundings spins and give rise to clusters of multi-spin correlations (shades) that grow in time. The second important element of the experiment is the possibility to collectively rotate the spins and perform a basis transformation that allows us to monitor the growth of multi-spin correlations by probing only the central spin~\cite{Ramanathan2003,Cho2005,Niknam2020}. 

We use the amplitudes of the multi-spin correlations to compute what we call correlation R\'enyi entropy. We find that after the saturation of the entanglement entropy, the correlation R\'enyi entropy keeps growing for times an order of magnitude longer, during which the larger clusters of multi-spin correlations build up. The experimental results show excellent agreement with our numerical simulations. We also perform a scaling analysis of the growth rate, the saturation value, and the equilibration time of the correlation R\'enyi entropy. Both the rate and the saturation point grow logarithmically as the size of the composite system increases, while the equilibration time is nearly independent of system size.


{\em Experimental System.--}   We work with a polycrystalline solid made of an ensemble of Triphenylphosphine molecules. Each molecule has a central \textsuperscript{31}P nuclear spin coupled to \textsuperscript{1}H spins via the heteronuclear dipolar interaction
\begin{equation} 
\H_{\text{CS-B}} =   \sum_{j}^{\text{N}} \omega_j  \sigmaz^{\text{CS}} \totimes \sigmaz^j \totimes \mathds{1}^{\tiny\totimes N-1} ,
\label{Eq:H}
\end{equation}
where `CS' stands for central spin, `B' for the finite bath with $N=15$ surrounding spins, and $\sigma_{\text{\tiny Z}}^j$ is the Pauli matrix for the  $ j^{\text{th}} $ spin. The coupling constants $\omega_j $ are determined by the orientation and the distance of the bath spins from the central spin, the majority having values below 1200 Hz (see distribution in~\cite{Niknam2020}). 

Quantum information resides initially in the central spin, which is prepared in a polarized state $ \rho^{\text{CS}}(0)= \sigma_{\text{\tiny X}}/2 $ ~\cite{NoteCS}, while the surrounding spins are in a fully mixed state  $\rho^{\text{B}} (0)=(\mathds{1}/2)^{\totimes N} $. The homonuclear dipolar interactions among the bath spins are averaged out by applying the MREV-8  pulse sequence, which cancels the interactions up to the third order of the Magnus expansion, including pulse imperfection effects~\cite{HaeberlenBook,Rhim73_mrev8,DybowskiBook}. During the entire time span of our experiment, the effects of external environments are also under control~\cite{Niknam2020}, so that the evolution of the density matrix of the composite system, $\rho(T) = U_{\text{CS-B}}(T) \rho(0) U_{\text{CS-B}}^{\dagger}(T)$, is effectively described by the unitary propagator $ U_{\text{CS-B}} (T)= \e^{-i \H_{\tiny\text{CS-B}} T}$. As the CS-B system evolves under the ZZ interaction, information from the central spin gets shared with the bath spins giving rise to clusters of multi-spin correlations.

\begin{figure}[htb]
 	\includegraphics*[scale=0.38]{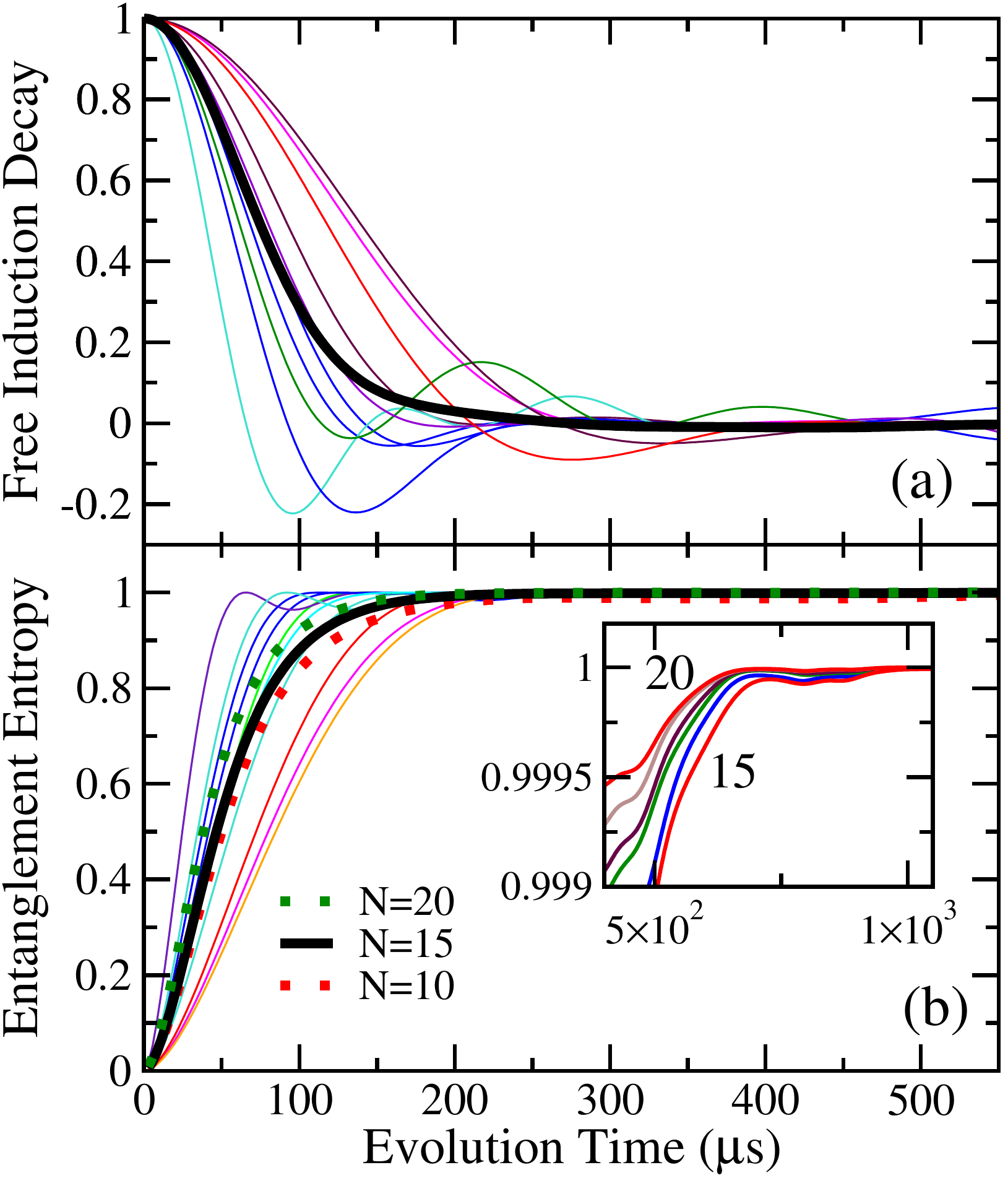}	   	 	 	 	 	 	
 	\caption{Free induction decay (a) and entanglement entropy (b) for a central spin coupled with 15 surrounding spins via the ZZ interaction in Eq.~(\ref{Eq:H}). Thin lines are obtained with random realizations of the coupling constants $\omega_j $, thick line gives the average over 300 orientations of the molecule. The dotted lines in (b) give the ensemble average for $N=10$ and $N=20$. Inset of panel (b): Averaged entangled entropy for $N=15, 16, \ldots, 20$ from bottom to top. }	
\label{fig02}
 \end{figure}

{\em FID and Entanglement Entropy.--}  The loss of information from the central spin can be quantified with the free induction decay (FID),
\begin{equation}
\text{FID} (T) \!=\! \Tr \{ \sigmax^{\text{cs}} \rho^{\text{cs}}(T) \} 
\!=\! \frac{1}{2^{N+1}} \!\! \sum_{k=1}^{2^{N+1}} \cos ( 2 \langle \phi_k| \H_{\text{CS-B}} |\phi_k\rangle t), 
\label{Eq:FID}
\end{equation}
where  $\rho^{\text{CS}}(T)= \Tr_{\text{B}} [\rho (T)]$ and $|\phi_k\rangle$ is one of the $2^{N+1}$ spin configurations in the $z$-direction, such as $|\uparrow \downarrow  \downarrow \ldots \uparrow \rangle$. The results from numerical simulations for $N=15$ are shown in Fig.~\ref{fig02}~(a). Thin lines correspond to representative random orientations of the molecules and the thick curve gives the average over 300 random realizations. The curve for the ensemble average is smooth and quickly saturates at $\overline{F} \sim 0$.

For our system,  the entanglement entropy between the central spin and the bath is simply a function of the FID,
\begin{eqnarray}
\label{Eq:ent}
S_{\text{ent}} (T) &=& - \Tr \left\{ \rho^{\text{CS}}(T) \text{log}_2 [ \rho^{\text{CS}}(T)  ] \right\} \\
&=& -[ f^+(T) \text{log}_2 f^+ (T)+ f^-(T) \text{log}_2 f^-(T)], \nonumber
\end{eqnarray}
where 
$f^{+,-}(T) = (1/2) \pm \text{FID}(T)$. In the main panel of Fig.~\ref{fig02}~(b), the thin lines show the entanglement entropy for representative random orientations of the molecules with $N=15$ and the thick curve gives the ensemble average. Similarly to the FID, $S_{\text{ent}} (T)$ evolves quickly and then saturates at the maximum entropy value $\overline{S}_{\text{ent}} \sim 1$. For both quantities, the saturation of the dynamics happens at $T\sim 200\mu$s. In Fig.~\ref{fig02}~(b),  we also show with dotted lines ensemble averages for $N=10$ and $N=20$. The slope of $S_{\text{ent}} (T)$ increases with system size, which suggests that the saturation should happen earlier for larger spin baths. In what follows, we compare these timescales with the saturation time obtained for the correlation R\'enyi entropy.


{\em Multi-Spin Correlations.--} During the evolution of the total density matrix,
\begin{eqnarray}
\label{Eq:rho}
\!\rho(T) \! &=& \!\e^{-i \H_{\tiny\text{CS-B}} T} \rho(0) \e^{i \H_{\tiny\text{CS-B}} T}\\
\!&\!=\!& \! \rho(0) \!+\!i T [\rho(0)\!,\!\H_{\text{CS-B}}]\!-\!\frac{T^2}{2}\![ [\rho(0)\!,\!\H_{\text{CS-B}}],\!\H_{\text{CS-B}}]+\!\ldots\!\nonumber \\
\!&\!=\!&\! \Cz_0(T) \sigmax^{\text{CS}} \totimes \mathds{1}^{\tiny\totimes N}  +
 \Cz_1(T) \sum_{j}^{N}\sigmay^{\text{CS}}  \totimes \sigmaz^j \totimes \mathds{1}^{\tiny\totimes N-1} \nonumber\\
&+& \Cz_2(T) \sum_{j\neq k}^{N}\sigmax^{\text{CS}}  \totimes \sigmaz^j \totimes \sigmaz^k \totimes \mathds{1}^{\tiny\totimes N-2}+ \cdots  \nonumber,
\end{eqnarray}
clusters of correlated spins build up and grow, each cluster having an amplitude $\Cz_m(T)$. In this picture, $m$ represents the number of bath spins that get correlated with the central spin, as determined by the number of bath spin operators $ \sigmaz $ that appears in each term of Eq.~(\ref{Eq:rho}). 

The uncorrelated term of amplitude $\Cz_0(T)$  is the only one that survives the partial trace used to obtain $\rho^{\text{CS}}(T)$ in Eq.~(\ref{Eq:FID}) and therefore the only one that contributes to $\text{FID}(T)$. This is also the case for the entanglement entropy, since both quantities are related. The decay of $\Cz_0(T)$ describes the loss of information from the central spin, which causes the decline of the observable NMR signal and the growth of the entanglement entropy. {\it However, to better understand the dynamics of the composite system, one needs a quantity that captures also the build-up of multi-spin correlations as determined by the higher orders terms with $m>0$}. 

To explain how we measure the higher order correlations, let us write the bath spin operators in terms of idempotent matrices $ {\cal I}_{\pm}=\frac{1}{2}(\mathds{1}\pm\sigma_{\text{\tiny Z}}) $~\cite{Somaroo:1998jy}. The first step in the production of correlations is to go from the term proportional to $\Cz_0(T)$ in Eq.~(\ref{Eq:rho}) to the term proportional to $\Cz_1(T)$, {\em i.e.} from $ \sigma_{\text{\tiny X}}^{\text{CS}}\totimes \mathds{1} =  \sigma_{\text{\tiny X}}^{\text{CS}} ({\cal I}_+^j  + {\cal I}_-^j )$, where there is no correlation, to 
\[
\label{corr_tran}
 \sigma_{\text{\tiny Y}}^{\text{CS}}\sigmaz^j =  \sigma_{\text{\tiny Y}}^{\text{CS}} ({\cal I}_+^j  - {\cal I}_-^j ) = \sigma_{\text{\tiny Y}}^{\text{CS}} {\cal I}_+^i  - \sigma_{\text{\tiny Y}}^{\text{CS}} {\cal I}_-^i,
\]
where there are two correlated terms, $\sigma_{\text{\tiny Y}}^{\text{CS}} {\cal I}_+^i $ with Hamming weight +1 and  $\sigma_{\text{\tiny Y}}^{\text{CS}} {\cal I}_-^i$ with Hamming weight -1. As the evolution further proceeds to 
$\sigmax^{\text{CS}} \sigmaz^j  \sigmaz^k = \sigma_{\text{\tiny X}}^{\text{CS}}  {\cal I}_+^j {\cal I}_+^k + \sigma_{\text{\tiny X}}^{\text{CS}} {\cal I}_-^j {\cal I}_-^k  - \sigma_{\text{\tiny X}}^{\text{CS}} {\cal I}_+^j {\cal I}_-^k - \sigma_{\text{\tiny X}}^{\text{CS}} {\cal I}_-^j {\cal I}_+^k,$
one gets four terms, one with Hamming weight + 2, one with -2, and two with Hamming weight zero. For $\sigmay^{\text{CS}}  \sigmaz^j \sigmaz^k \sigmaz^l$, there are 8 terms, one with Hamming weight +3, one -3, three +1, and three -1. And like this successively, each step presenting a binomial distribution of Hamming weights.

Our experiment is designed to directly measure the amplitudes $\Cx_n(T)$ of all terms that have a given Hamming weight $n$. Clearly $\Cx_n(T) = \Cx_{-n}(T)$ and $\Cx_n(T)\neq \Cz_m(T)$. The explanation for the superscript  x  in $\Cx_n(T)$ is given below.

The multi-spin correlations are found in the state of the composite system, while the NMR experiment probes only the central spin. To access the Hamming weights amplitudes, we then extend to our heteronuclear system, a basis-change technique that has been used to study multiple quantum coherences in homonuclear systems~\cite{Baum1985,Ramanathan2003,Cho2005}. The strategy consists in applying to the bath spins at time $T$ a collective rotation along the $x$-axis, that is described by the operator $R_x(\phi)=\exp\left( i \frac{\phi}{2} \sum_j \mathds{1}^{\text{CS}} \totimes \mathds{1}^1 \totimes \cdots \totimes \sigma_{\text{\tiny X}}^j\totimes \cdots\totimes \mathds{1}^N  \right)$. This way, each Hamming weight $n$ gets encoded in the phase factor $e^{i n \phi} $  according to
\begin{equation} \label{eq:phaseencoding}
\rho_{\phi}(T)=R_x(\phi) \rho(T) R_x^{\dagger}(\phi)=\sum_{n} e^{i n \phi} \Cx_n(T) \rho_{n}^{\text x}  ,
\end{equation}
The amplitudes $\Cx_n(T)$ of the multi-spin correlation orders in the $x$-quantization axis are identical to the amplitudes of all correlation terms characterized by the Hamming weight orders of  ${\cal I}_{\pm} $ and denoted by $ \rho_{n}^{\text x} $. 

Next, we apply a $ \pi $-pulse to the central spin to reverse the CS-B dynamics, 
$ \rho_{\phi}(2T) \!=\! U_{\text{CS-B}}^{\dagger}(T)\rho_{\phi}(T)U_{\text{CS-B}}(T)$,
and create the observable NMR signal at time $ 2T $, which is the overlap between the initial and final states of the central spin,
$\text{SIG}_{\phi}(2T) =\Tr [ \Tr_B [\rho_{\phi} (2T)].  \sigma^{\text{\tiny CS}}_{\text{\tiny X}}   ]$.
The encoding phase factors $ e^{i n \phi}  $ are now contained in the observable signal. They are recorded for increments of rotation angle $ \phi \in [0,2\pi] $. By performing a Fourier transform to the array of observed signals with various values of $ \phi $, we obtain the squared-amplitudes (intensities) of the Hamming weights, $ \Cx_n(T)^2$, where $\sum_{n=-N}^N \Cx_n(T)^2=1$.

\begin{figure}[h!]
	\includegraphics*[scale=0.33]{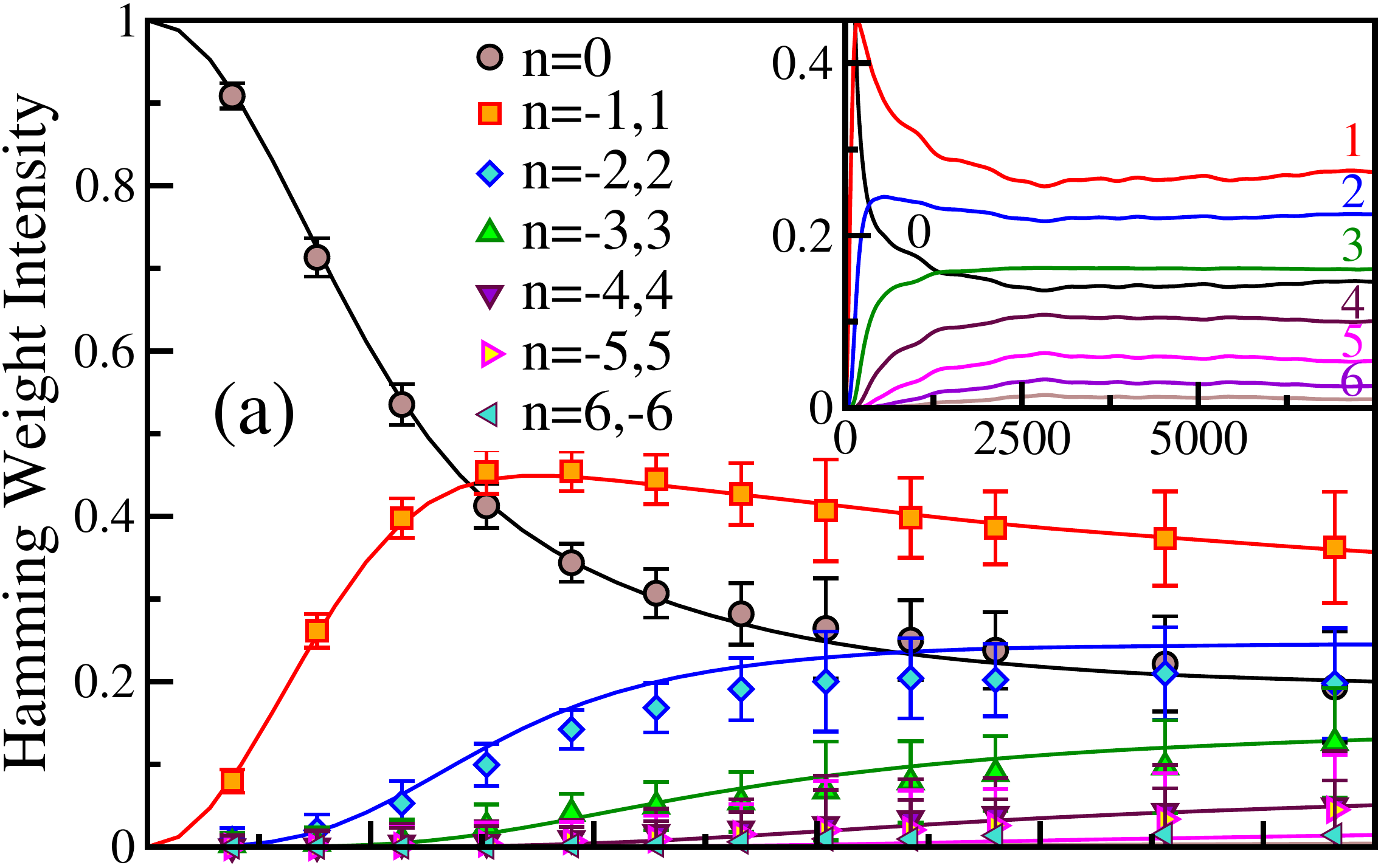}
	\includegraphics*[scale=0.33]{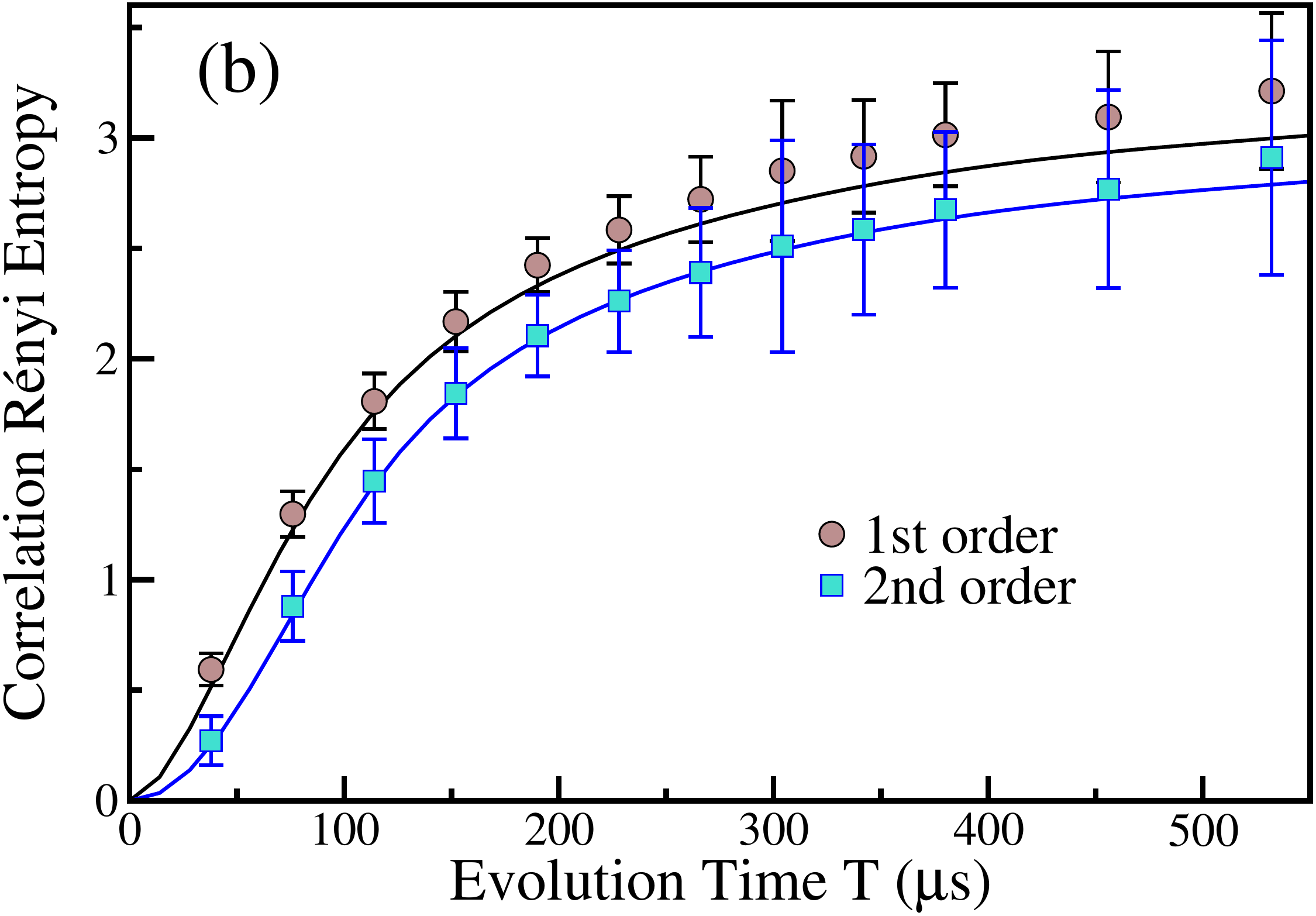}		 	 	 	
	\caption{Evolution of the intensities of Hamming weights $\Cx_n(T)^2$ in (a) and for longer time in the inset,  and the first and second order correlation R\'enyi entropies in (b). Symbols represent the experimental data and solid lines give the numerical results for $N=15$ averaged over 300 orientations of the molecule. }
	\label{fig03}
\end{figure}

The evolution of the Hamming weights intensities is shown in Fig.~\ref{fig03}~(a).  The agreement between the experimental data and the numerical simulations for orders up to $n=-6,6$ is excellent. Higher order terms develop at even longer times and are more challenging to detect experimentally. This figure reveals the details of how information lost from the central spin gets shared with the surrounding qubits.

{\em Correlation R\'enyi Entropies.--}  We use the Hamming weights intensities $\Cx_n(T)^2$  to compute the correlation R\'enyi entropy. The first and second order correlation R\'enyi entropies are respectively defined as
\begin{eqnarray}
S_1 &=& - \sum_n \Cx_n(T)^2 \text{log}_2 \Cx_n(T)^2  , \\
S_2 &=& - \text{log}_2 \left( \sum_n \Cx_n(T)^4 \right) .
\end{eqnarray}
They describe the growth of multi-spin correlations in the $x$-axis. Absence of correlations implies that $S_1^{\text{min}}=S_2^{\text{min}} =0$, while the homogeneous distribution of correlations among all orders, that is $\Cx_n(T)^2 = (2N+1)^{-1}$, leads to the maximum value, $S_1^{\text{max}}=S_2^{\text{max}} = \text{log}_2 (2N+1)$. 

\begin{figure}[hbt]
	\includegraphics*[scale=0.35]{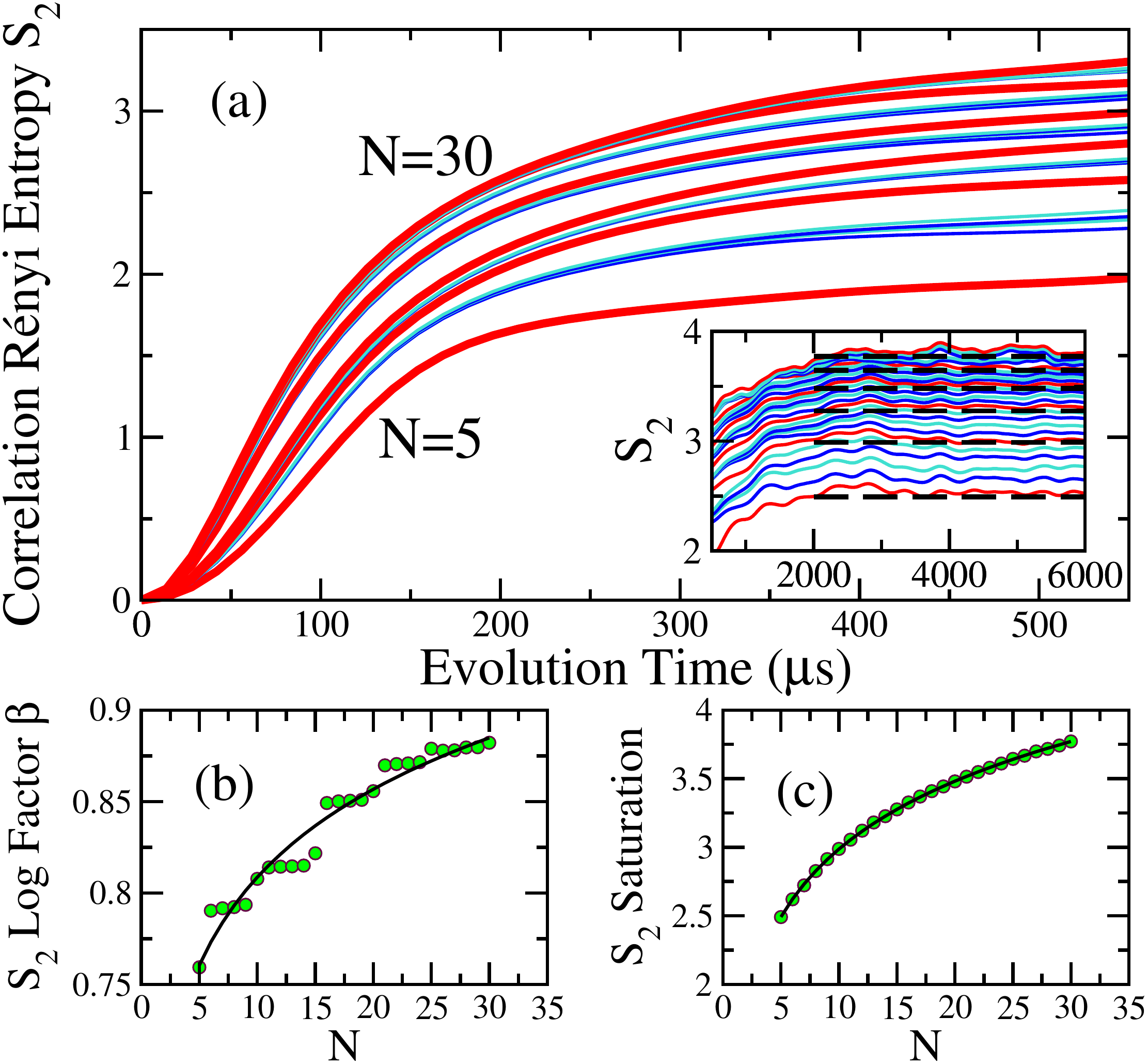}	  	 	 	 	
	\caption{In (a): Evolution of the correlation R\'enyi entropy $S_2$ for $N=5, \ldots 30$ averaged over 300 orientations of the molecules for $N=5, \ldots 24$ and over 100 or 20 realizations for the largest sizes. The thick red lines correspond to bath sizes that are multiples of 5. The inset in (a) shows the curves up to saturation and $\overline{S_2}$ is indicated with dashed lines. In (b): scaling analysis of the factor $\beta$ in $S_2(T) \sim -4.18 + \beta \log_2(T)$ (symbols) fitted with $0.65 + 0.07 \ln(N)$ (solid line). In (c): scaling analysis of the saturation values of $S_2$ (symbols) fitted with $1.34 + 0.71 \ln(N)$ (solid line).
	} 
	\label{fig04}
\end{figure}

The experimental data for both entropies are compared with numerical simulations in Fig.~\ref{fig03}~(b). One sees that the growth of $S_{1,2}$ is not complete during the timescale of our experiment. The correlation R\'enyi entropy keeps increasing for $T> 500\mu$s, implying that the growth of the volume of correlations has not yet ceased and correlations of higher orders are still developing. In fact, as the simulations for different bath sizes in the inset of Fig.~\ref{fig04}~(a) indicate, saturation happens at $T\sim 2000\mu$s. This contrasts with the entanglement entropy displayed in Fig.~\ref{fig02}~(b), where the curves are already flat for times one order of magnitude shorter, at $T\sim 200\mu$s. 

The discrepancy between the timescales for the equilibration of $S_{1,2}$ and $S_{\text{ent}}$ motivated us to have a closer look at the saturation of the entanglement entropy. By significantly zooming in the $y$-axis of Fig.~\ref{fig02}~(b), we observe in the inset that $S_{\text{ent}}$ for different bath sizes actually keeps increasing for $T> 500\mu$s. It is only because we have a detailed picture of the growth of the volume of correlations, that we could have expected the existence of this residual increase. The evolution of the entanglement entropy, just as the FID, reflects the loss of information from the central spin, as characterized by the decay of $\Cz_0(T)$, and this decay happens simultaneously with the growth of the higher order correlations. While the necessary precision to detect the growth of $S_{\text{ent}}$ for $T> 200\mu$s is experimentally unreachable, the experimental increase of $S_{1,2}$ at these long timescales is evident in Fig.~\ref{fig03}~(b).

{\em Equilibration.--} The complete saturation of the correlation R\'enyi entropy takes place once the clusters of correlated spins stop growing, that is, when the Hamming weights intensities become constant, as seen in the inset of Fig.~\ref{fig03}~(a). To estimate the timescale for the equilibration and how it depends on the bath size, we study numerically in Fig.~\ref{fig04}~(a) the evolution of $S_2$  for baths ranging from $N=5$ to $N=30$. 

We compute the saturation value of the entropy, $\overline{S_2}$, by averaging the values of $S_2(T)$ for $T>5000\mu$s, when the curves are clearly flat, as seen in the inset of Fig.~\ref{fig04}~(a). We then obtain the equilibration times $T_{eq}$ by verifying where each numerical curve of $S_2(T)$ first crosses its saturation point. We find that $T_{eq} = (2052.1 \pm 163.7)\mu$s for $N=5 \ldots 30$, while considering only larger bath sizes, $N\geq 15$, the fluctuations decrease and we get $T_{eq} = (2145.5 \pm 39.4) \mu$s.

The fact that $T_{eq}$ is nearly independent of system size is likely a consequence of the fact that both the growth rate of $S_2(T)$ and also its saturation value scale as $\ln(N)$. We fit the evolution of $S_2$ in the interval $50\mu$s $<T<300\mu$s with the logarithmic function $\alpha+ \beta \log_2(x)$, where $\alpha$ and $\beta$ are fitting constants. The fitting improves for larger system sizes and we find that the factor $\beta$ increases as $\ln(N)$, as shown in Fig.~\ref{fig04}~(b). Since the bath is simulated using rings of 5 spins, similar to the physical sample in Fig.~\ref{fig01}, whenever a new ring is introduced, $\beta$ surges with the addition of the first spins, which are closer and thus more strongly coupled with the central spin. As for $\overline{S_2}$, the scaling analysis in Fig.~\ref{fig04}~(c) demonstrates that it also increases as $\ln(N)$.

{\em Conclusion.--} We introduced and experimentally measured the correlation R\'enyi entropy, which quantifies the volume of multi-spin correlations. The experimental resources needed to measure this entropy in the central spin model scales linearly with the size of the composite system. While the entanglement entropy, $S_{\text{ent}}$, quantifies the loss of information from the central spin, the correlation R\'enyi entropy, $S_{\text{1,2}}$, provides a more detailed picture of the dynamics of the composite system by capturing how that information gets shared among the bath spins. Most notably, the $S_{\text{1,2}}$ saturates at a time that is an order of magnitude larger than the saturation time for $S_{\text{ent}}$.  The correlation R\'enyi entropy opens interesting perspectives for experimental detection of many-body correlations growth and the spread of quantum information in quantum devices. It may find applications in quantum error correction codes, where information is encoded in long-range quantum many-body states, and in studies of the propagation speed of correlations.

\begin{acknowledgments}
MN and LFS thank the hospitality of the Simons Center for Geometry and Physics at Stony Brook University, where some of the research for this paper was performed. LFS is supported by the NSF Grant No. DMR-1936006. The research results communicated here would not be possible without the significant contributions of the Canada First Research Excellence Fund.
\end{acknowledgments}


\begin{thebibliography}{35}%
\makeatletter
\providecommand \@ifxundefined [1]{%
 \@ifx{#1\undefined}
}%
\providecommand \@ifnum [1]{%
 \ifnum #1\expandafter \@firstoftwo
 \else \expandafter \@secondoftwo
 \fi
}%
\providecommand \@ifx [1]{%
 \ifx #1\expandafter \@firstoftwo
 \else \expandafter \@secondoftwo
 \fi
}%
\providecommand \natexlab [1]{#1}%
\providecommand \enquote  [1]{``#1''}%
\providecommand \bibnamefont  [1]{#1}%
\providecommand \bibfnamefont [1]{#1}%
\providecommand \citenamefont [1]{#1}%
\providecommand \href@noop [0]{\@secondoftwo}%
\providecommand \href [0]{\begingroup \@sanitize@url \@href}%
\providecommand \@href[1]{\@@startlink{#1}\@@href}%
\providecommand \@@href[1]{\endgroup#1\@@endlink}%
\providecommand \@sanitize@url [0]{\catcode `\\12\catcode `\$12\catcode
  `\&12\catcode `\#12\catcode `\^12\catcode `\_12\catcode `\%12\relax}%
\providecommand \@@startlink[1]{}%
\providecommand \@@endlink[0]{}%
\providecommand \url  [0]{\begingroup\@sanitize@url \@url }%
\providecommand \@url [1]{\endgroup\@href {#1}{\urlprefix }}%
\providecommand \urlprefix  [0]{URL }%
\providecommand \Eprint [0]{\href }%
\providecommand \doibase [0]{http://dx.doi.org/}%
\providecommand \selectlanguage [0]{\@gobble}%
\providecommand \bibinfo  [0]{\@secondoftwo}%
\providecommand \bibfield  [0]{\@secondoftwo}%
\providecommand \translation [1]{[#1]}%
\providecommand \BibitemOpen [0]{}%
\providecommand \bibitemStop [0]{}%
\providecommand \bibitemNoStop [0]{.\EOS\space}%
\providecommand \EOS [0]{\spacefactor3000\relax}%
\providecommand \BibitemShut  [1]{\csname bibitem#1\endcsname}%
\let\auto@bib@innerbib\@empty
\bibitem [{\citenamefont {Neill}\ \emph {et~al.}(2016)\citenamefont {Neill},
  \citenamefont {Roushan}, \citenamefont {Fang}, \citenamefont {Chen},
  \citenamefont {Kolodrubetz}, \citenamefont {Chen}, \citenamefont {Megrant},
  \citenamefont {Barends}, \citenamefont {Campbell}, \citenamefont {Chiaro},
  \citenamefont {Dunsworth}, \citenamefont {Jeffrey}, \citenamefont {Kelly},
  \citenamefont {Mutus}, \citenamefont {O'Malley}, \citenamefont {Quintana},
  \citenamefont {Sank}, \citenamefont {Vainsencher}, \citenamefont {Wenner},
  \citenamefont {White}, \citenamefont {Polkovnikov},\ and\ \citenamefont
  {Martinis}}]{Neill2016}%
  \BibitemOpen
  \bibfield  {author} {\bibinfo {author} {\bibfnamefont {C.}~\bibnamefont
  {Neill}}, \bibinfo {author} {\bibfnamefont {P.}~\bibnamefont {Roushan}},
  \bibinfo {author} {\bibfnamefont {M.}~\bibnamefont {Fang}}, \bibinfo {author}
  {\bibfnamefont {Y.}~\bibnamefont {Chen}}, \bibinfo {author} {\bibfnamefont
  {M.}~\bibnamefont {Kolodrubetz}}, \bibinfo {author} {\bibfnamefont
  {Z.}~\bibnamefont {Chen}}, \bibinfo {author} {\bibfnamefont {A.}~\bibnamefont
  {Megrant}}, \bibinfo {author} {\bibfnamefont {R.}~\bibnamefont {Barends}},
  \bibinfo {author} {\bibfnamefont {B.}~\bibnamefont {Campbell}}, \bibinfo
  {author} {\bibfnamefont {B.}~\bibnamefont {Chiaro}}, \bibinfo {author}
  {\bibfnamefont {A.}~\bibnamefont {Dunsworth}}, \bibinfo {author}
  {\bibfnamefont {E.}~\bibnamefont {Jeffrey}}, \bibinfo {author} {\bibfnamefont
  {J.}~\bibnamefont {Kelly}}, \bibinfo {author} {\bibfnamefont
  {J.}~\bibnamefont {Mutus}}, \bibinfo {author} {\bibfnamefont {P.~J.~J.}\
  \bibnamefont {O'Malley}}, \bibinfo {author} {\bibfnamefont {C.}~\bibnamefont
  {Quintana}}, \bibinfo {author} {\bibfnamefont {D.}~\bibnamefont {Sank}},
  \bibinfo {author} {\bibfnamefont {A.}~\bibnamefont {Vainsencher}}, \bibinfo
  {author} {\bibfnamefont {J.}~\bibnamefont {Wenner}}, \bibinfo {author}
  {\bibfnamefont {T.~C.}\ \bibnamefont {White}}, \bibinfo {author}
  {\bibfnamefont {A.}~\bibnamefont {Polkovnikov}}, \ and\ \bibinfo {author}
  {\bibfnamefont {J.~M.}\ \bibnamefont {Martinis}},\ }\bibfield  {title}
  {\enquote {\bibinfo {title} {Ergodic dynamics and thermalization in an
  isolated quantum system},}\ }\href {http://dx.doi.org/10.1038/nphys3830}
  {\bibfield  {journal} {\bibinfo  {journal} {Nat. Phys.}\ }\textbf {\bibinfo
  {volume} {12}},\ \bibinfo {pages} {1037} (\bibinfo {year}
  {2016})}\BibitemShut {NoStop}%
\bibitem [{\citenamefont {Kaufman}\ \emph {et~al.}(2016)\citenamefont
  {Kaufman}, \citenamefont {M.~Eric~Tai}, \citenamefont {Rispoli},
  \citenamefont {Schittko}, \citenamefont {Preiss},\ and\ \citenamefont
  {Greiner}}]{Kaufman2016}%
  \BibitemOpen
  \bibfield  {author} {\bibinfo {author} {\bibfnamefont {Adam~M.}\ \bibnamefont
  {Kaufman}}, \bibinfo {author} {\bibfnamefont {Alexander~Lukin}\ \bibnamefont
  {M.~Eric~Tai}}, \bibinfo {author} {\bibfnamefont {Matthew}\ \bibnamefont
  {Rispoli}}, \bibinfo {author} {\bibfnamefont {Robert}\ \bibnamefont
  {Schittko}}, \bibinfo {author} {\bibfnamefont {Philipp~M.}\ \bibnamefont
  {Preiss}}, \ and\ \bibinfo {author} {\bibfnamefont {Markus}\ \bibnamefont
  {Greiner}},\ }\bibfield  {title} {\enquote {\bibinfo {title} {Quantum
  thermalization through entanglement in an isolated many-body system},}\
  }\href {\doibase 10.1126/science.aaf6725} {\bibfield  {journal} {\bibinfo
  {journal} {Science}\ }\textbf {\bibinfo {volume} {353}},\ \bibinfo {pages}
  {794} (\bibinfo {year} {2016})}\BibitemShut {NoStop}%
\bibitem [{\citenamefont {Brydges}\ \emph {et~al.}(2019)\citenamefont
  {Brydges}, \citenamefont {Elben}, \citenamefont {Jurcevic}, \citenamefont
  {Vermersch}, \citenamefont {Maier}, \citenamefont {Lanyon}, \citenamefont
  {Zoller}, \citenamefont {Blatt},\ and\ \citenamefont {Roos}}]{Brydges2019}%
  \BibitemOpen
  \bibfield  {author} {\bibinfo {author} {\bibfnamefont {Tiff}\ \bibnamefont
  {Brydges}}, \bibinfo {author} {\bibfnamefont {Andreas}\ \bibnamefont
  {Elben}}, \bibinfo {author} {\bibfnamefont {Petar}\ \bibnamefont {Jurcevic}},
  \bibinfo {author} {\bibfnamefont {Beno{\^\i}t}\ \bibnamefont {Vermersch}},
  \bibinfo {author} {\bibfnamefont {Christine}\ \bibnamefont {Maier}}, \bibinfo
  {author} {\bibfnamefont {Ben~P.}\ \bibnamefont {Lanyon}}, \bibinfo {author}
  {\bibfnamefont {Peter}\ \bibnamefont {Zoller}}, \bibinfo {author}
  {\bibfnamefont {Rainer}\ \bibnamefont {Blatt}}, \ and\ \bibinfo {author}
  {\bibfnamefont {Christian~F.}\ \bibnamefont {Roos}},\ }\bibfield  {title}
  {\enquote {\bibinfo {title} {Probing {R}{\'e}nyi entanglement entropy via
  randomized measurements},}\ }\href {\doibase 10.1126/science.aau4963}
  {\bibfield  {journal} {\bibinfo  {journal} {Science}\ }\textbf {\bibinfo
  {volume} {364}},\ \bibinfo {pages} {260--263} (\bibinfo {year}
  {2019})}\BibitemShut {NoStop}%
\bibitem [{\citenamefont {Lu}\ \emph {et~al.}(2010)\citenamefont {Lu},
  \citenamefont {Wang},\ and\ \citenamefont {Sun}}]{Lu2010}%
  \BibitemOpen
  \bibfield  {author} {\bibinfo {author} {\bibfnamefont {Xiao-Ming}\
  \bibnamefont {Lu}}, \bibinfo {author} {\bibfnamefont {Xiaoguang}\
  \bibnamefont {Wang}}, \ and\ \bibinfo {author} {\bibfnamefont {C.~P.}\
  \bibnamefont {Sun}},\ }\bibfield  {title} {\enquote {\bibinfo {title}
  {Quantum fisher information flow and non-markovian processes of open
  systems},}\ }\href {\doibase 10.1103/PhysRevA.82.042103} {\bibfield
  {journal} {\bibinfo  {journal} {Phys. Rev. A}\ }\textbf {\bibinfo {volume}
  {82}},\ \bibinfo {pages} {042103} (\bibinfo {year} {2010})}\BibitemShut
  {NoStop}%
\bibitem [{\citenamefont {G\"arttner}\ \emph {et~al.}(2018)\citenamefont
  {G\"arttner}, \citenamefont {Hauke},\ and\ \citenamefont
  {Rey}}]{Garttner2018}%
  \BibitemOpen
  \bibfield  {author} {\bibinfo {author} {\bibfnamefont {Martin}\ \bibnamefont
  {G\"arttner}}, \bibinfo {author} {\bibfnamefont {Philipp}\ \bibnamefont
  {Hauke}}, \ and\ \bibinfo {author} {\bibfnamefont {Ana~Maria}\ \bibnamefont
  {Rey}},\ }\bibfield  {title} {\enquote {\bibinfo {title} {Relating
  out-of-time-order correlations to entanglement via multiple-quantum
  coherences},}\ }\href {\doibase 10.1103/PhysRevLett.120.040402} {\bibfield
  {journal} {\bibinfo  {journal} {Phys. Rev. Lett.}\ }\textbf {\bibinfo
  {volume} {120}},\ \bibinfo {pages} {040402} (\bibinfo {year}
  {2018})}\BibitemShut {NoStop}%
\bibitem [{\citenamefont {Smith}\ \emph {et~al.}(2016)\citenamefont {Smith},
  \citenamefont {Lee}, \citenamefont {Richerme}, \citenamefont {Neyenhuis},
  \citenamefont {Hess}, \citenamefont {Hauke}, \citenamefont {Heyl},
  \citenamefont {Huse},\ and\ \citenamefont {Monroe}}]{Smith2015}%
  \BibitemOpen
  \bibfield  {author} {\bibinfo {author} {\bibfnamefont {J.}~\bibnamefont
  {Smith}}, \bibinfo {author} {\bibfnamefont {A.}~\bibnamefont {Lee}}, \bibinfo
  {author} {\bibfnamefont {P.}~\bibnamefont {Richerme}}, \bibinfo {author}
  {\bibfnamefont {B.}~\bibnamefont {Neyenhuis}}, \bibinfo {author}
  {\bibfnamefont {P.~W.}\ \bibnamefont {Hess}}, \bibinfo {author}
  {\bibfnamefont {P.}~\bibnamefont {Hauke}}, \bibinfo {author} {\bibfnamefont
  {M.}~\bibnamefont {Heyl}}, \bibinfo {author} {\bibfnamefont {D.~A.}\
  \bibnamefont {Huse}}, \ and\ \bibinfo {author} {\bibfnamefont
  {C.}~\bibnamefont {Monroe}},\ }\bibfield  {title} {\enquote {\bibinfo {title}
  {Many-body localization in a quantum simulator with programmable random
  disorder},}\ }\href {\doibase 10.1038/nphys3783} {\bibfield  {journal}
  {\bibinfo  {journal} {Nat. Phys.}\ }\textbf {\bibinfo {volume} {12}},\
  \bibinfo {pages} {907} (\bibinfo {year} {2016})}\BibitemShut {NoStop}%
\bibitem [{\citenamefont {Niknam}\ \emph {et~al.}(2020)\citenamefont {Niknam},
  \citenamefont {Santos},\ and\ \citenamefont {Cory}}]{Niknam2020}%
  \BibitemOpen
  \bibfield  {author} {\bibinfo {author} {\bibfnamefont {Mohamad}\ \bibnamefont
  {Niknam}}, \bibinfo {author} {\bibfnamefont {Lea~F.}\ \bibnamefont {Santos}},
  \ and\ \bibinfo {author} {\bibfnamefont {David~G.}\ \bibnamefont {Cory}},\
  }\bibfield  {title} {\enquote {\bibinfo {title} {Sensitivity of quantum
  information to environment perturbations measured with a nonlocal
  out-of-time-order correlation function},}\ }\href {\doibase
  10.1103/PhysRevResearch.2.013200} {\bibfield  {journal} {\bibinfo  {journal}
  {Phys. Rev. Res.}\ }\textbf {\bibinfo {volume} {2}},\ \bibinfo {pages}
  {013200} (\bibinfo {year} {2020})}\BibitemShut {NoStop}%
\bibitem [{\citenamefont {Flambaum}\ and\ \citenamefont
  {Izrailev}(2001)}]{Flambaum2001b}%
  \BibitemOpen
  \bibfield  {author} {\bibinfo {author} {\bibfnamefont {V.~V.}\ \bibnamefont
  {Flambaum}}\ and\ \bibinfo {author} {\bibfnamefont {F.~M.}\ \bibnamefont
  {Izrailev}},\ }\bibfield  {title} {\enquote {\bibinfo {title} {Entropy
  production and wave packet dynamics in the fock space of closed chaotic
  many-body systems},}\ }\href@noop {} {\bibfield  {journal} {\bibinfo
  {journal} {Phys. Rev. E}\ }\textbf {\bibinfo {volume} {64}},\ \bibinfo
  {pages} {036220} (\bibinfo {year} {2001})}\BibitemShut {NoStop}%
\bibitem [{\citenamefont {Asplund}\ and\ \citenamefont
  {Bernamonti}(2014)}]{Asplund2014}%
  \BibitemOpen
  \bibfield  {author} {\bibinfo {author} {\bibfnamefont {Curtis~T.}\
  \bibnamefont {Asplund}}\ and\ \bibinfo {author} {\bibfnamefont {Alice}\
  \bibnamefont {Bernamonti}},\ }\bibfield  {title} {\enquote {\bibinfo {title}
  {Mutual information after a local quench in conformal field theory},}\ }\href
  {\doibase 10.1103/PhysRevD.89.066015} {\bibfield  {journal} {\bibinfo
  {journal} {Phys. Rev. D}\ }\textbf {\bibinfo {volume} {89}},\ \bibinfo
  {pages} {066015} (\bibinfo {year} {2014})}\BibitemShut {NoStop}%
\bibitem [{\citenamefont {Bianchi}\ \emph {et~al.}(2018)\citenamefont
  {Bianchi}, \citenamefont {Hackl},\ and\ \citenamefont
  {Yokomizo}}]{Bianchi2018}%
  \BibitemOpen
  \bibfield  {author} {\bibinfo {author} {\bibfnamefont {Eugenio}\ \bibnamefont
  {Bianchi}}, \bibinfo {author} {\bibfnamefont {Lucas}\ \bibnamefont {Hackl}},
  \ and\ \bibinfo {author} {\bibfnamefont {Nelson}\ \bibnamefont {Yokomizo}},\
  }\bibfield  {title} {\enquote {\bibinfo {title} {Linear growth of the
  entanglement entropy and the kolmogorov-sinai rate},}\ }\href@noop {}
  {\bibfield  {journal} {\bibinfo  {journal} {J. High Energy Phys.}\ }\textbf
  {\bibinfo {volume} {2018}},\ \bibinfo {pages} {25} (\bibinfo {year}
  {2018})}\BibitemShut {NoStop}%
\bibitem [{\citenamefont {Santos}\ \emph
  {et~al.}(2012{\natexlab{a}})\citenamefont {Santos}, \citenamefont
  {Borgonovi},\ and\ \citenamefont {Izrailev}}]{Santos2012PRL}%
  \BibitemOpen
  \bibfield  {author} {\bibinfo {author} {\bibfnamefont {L.~F.}\ \bibnamefont
  {Santos}}, \bibinfo {author} {\bibfnamefont {F.}~\bibnamefont {Borgonovi}}, \
  and\ \bibinfo {author} {\bibfnamefont {F.~M.}\ \bibnamefont {Izrailev}},\
  }\bibfield  {title} {\enquote {\bibinfo {title} {Chaos and statistical
  relaxation in quantum systems of interacting particles},}\ }\href@noop {}
  {\bibfield  {journal} {\bibinfo  {journal} {Phys. Rev. Lett.}\ }\textbf
  {\bibinfo {volume} {108}},\ \bibinfo {pages} {094102} (\bibinfo {year}
  {2012}{\natexlab{a}})}\BibitemShut {NoStop}%
\bibitem [{\citenamefont {Vidmar}\ and\ \citenamefont
  {Rigol}(2017)}]{Vidmar2017b}%
  \BibitemOpen
  \bibfield  {author} {\bibinfo {author} {\bibfnamefont {Lev}\ \bibnamefont
  {Vidmar}}\ and\ \bibinfo {author} {\bibfnamefont {Marcos}\ \bibnamefont
  {Rigol}},\ }\bibfield  {title} {\enquote {\bibinfo {title} {Entanglement
  entropy of eigenstates of quantum chaotic Hamiltonians},}\ }\href {\doibase
  10.1103/PhysRevLett.119.220603} {\bibfield  {journal} {\bibinfo  {journal}
  {Phys. Rev. Lett.}\ }\textbf {\bibinfo {volume} {119}},\ \bibinfo {pages}
  {220603} (\bibinfo {year} {2017})}\BibitemShut {NoStop}%
\bibitem [{\citenamefont {Borgonovi}\ \emph {et~al.}(2019)\citenamefont
  {Borgonovi}, \citenamefont {Izrailev},\ and\ \citenamefont
  {Santos}}]{Borgonovi2019R}%
  \BibitemOpen
  \bibfield  {author} {\bibinfo {author} {\bibfnamefont {Fausto}\ \bibnamefont
  {Borgonovi}}, \bibinfo {author} {\bibfnamefont {Felix~M.}\ \bibnamefont
  {Izrailev}}, \ and\ \bibinfo {author} {\bibfnamefont {Lea~F.}\ \bibnamefont
  {Santos}},\ }\bibfield  {title} {\enquote {\bibinfo {title} {Exponentially
  fast dynamics of chaotic many-body systems},}\ }\href {\doibase
  10.1103/PhysRevE.99.010101} {\bibfield  {journal} {\bibinfo  {journal} {Phys.
  Rev. E}\ }\textbf {\bibinfo {volume} {99}},\ \bibinfo {pages} {010101}
  (\bibinfo {year} {2019})}\BibitemShut {NoStop}%
\bibitem [{\citenamefont {\ifmmode \check{Z}\else
  \v{Z}\fi{}nidari\ifmmode~\check{c}\else \v{c}\fi{}}\ \emph
  {et~al.}(2008)\citenamefont {\ifmmode \check{Z}\else
  \v{Z}\fi{}nidari\ifmmode~\check{c}\else \v{c}\fi{}}, \citenamefont {Prosen},\
  and\ \citenamefont {Prelov\ifmmode~\check{s}\else
  \v{s}\fi{}ek}}]{Znidaric2008}%
  \BibitemOpen
  \bibfield  {author} {\bibinfo {author} {\bibfnamefont {Marko}\ \bibnamefont
  {\ifmmode \check{Z}\else \v{Z}\fi{}nidari\ifmmode~\check{c}\else
  \v{c}\fi{}}}, \bibinfo {author} {\bibfnamefont {T.}~\bibnamefont {Prosen}}, \
  and\ \bibinfo {author} {\bibfnamefont {Peter}\ \bibnamefont
  {Prelov\ifmmode~\check{s}\else \v{s}\fi{}ek}},\ }\bibfield  {title} {\enquote
  {\bibinfo {title} {Many-body localization in the {H}eisenberg {XXZ} magnet in
  a random field},}\ }\href {\doibase 10.1103/PhysRevB.77.064426} {\bibfield
  {journal} {\bibinfo  {journal} {Phys. Rev. B}\ }\textbf {\bibinfo {volume}
  {77}},\ \bibinfo {pages} {064426} (\bibinfo {year} {2008})}\BibitemShut
  {NoStop}%
\bibitem [{\citenamefont {Bardarson}\ \emph {et~al.}(2012)\citenamefont
  {Bardarson}, \citenamefont {Pollmann},\ and\ \citenamefont
  {Moore}}]{Bardarson2012}%
  \BibitemOpen
  \bibfield  {author} {\bibinfo {author} {\bibfnamefont {Jens~H.}\ \bibnamefont
  {Bardarson}}, \bibinfo {author} {\bibfnamefont {Frank}\ \bibnamefont
  {Pollmann}}, \ and\ \bibinfo {author} {\bibfnamefont {Joel~E.}\ \bibnamefont
  {Moore}},\ }\bibfield  {title} {\enquote {\bibinfo {title} {Unbounded growth
  of entanglement in models of many-body localization},}\ }\href {\doibase
  10.1103/PhysRevLett.109.017202} {\bibfield  {journal} {\bibinfo  {journal}
  {Phys. Rev. Lett.}\ }\textbf {\bibinfo {volume} {109}},\ \bibinfo {pages}
  {017202} (\bibinfo {year} {2012})}\BibitemShut {NoStop}%
\bibitem [{\citenamefont {Torres-Herrera}\ and\ \citenamefont
  {Santos}(2017)}]{Torres2017}%
  \BibitemOpen
  \bibfield  {author} {\bibinfo {author} {\bibfnamefont {E.~J.}\ \bibnamefont
  {Torres-Herrera}}\ and\ \bibinfo {author} {\bibfnamefont {Lea~F.}\
  \bibnamefont {Santos}},\ }\bibfield  {title} {\enquote {\bibinfo {title}
  {Extended nonergodic states in disordered many-body quantum systems},}\
  }\href {\doibase 10.1002/andp.201600284} {\bibfield  {journal} {\bibinfo
  {journal} {Ann. Phys. (Berlin)}\ }\textbf {\bibinfo {volume} {529}},\
  \bibinfo {pages} {1600284} (\bibinfo {year} {2017})}\BibitemShut {NoStop}%
\bibitem [{\citenamefont {Lukin}\ \emph {et~al.}(2019)\citenamefont {Lukin},
  \citenamefont {Rispoli}, \citenamefont {Schittko}, \citenamefont {Tai},
  \citenamefont {Kaufman}, \citenamefont {Choi}, \citenamefont {Khemani},
  \citenamefont {L{\'e}onard},\ and\ \citenamefont {Greiner}}]{Lukin2019}%
  \BibitemOpen
  \bibfield  {author} {\bibinfo {author} {\bibfnamefont {Alexander}\
  \bibnamefont {Lukin}}, \bibinfo {author} {\bibfnamefont {Matthew}\
  \bibnamefont {Rispoli}}, \bibinfo {author} {\bibfnamefont {Robert}\
  \bibnamefont {Schittko}}, \bibinfo {author} {\bibfnamefont {M.~Eric}\
  \bibnamefont {Tai}}, \bibinfo {author} {\bibfnamefont {Adam~M.}\ \bibnamefont
  {Kaufman}}, \bibinfo {author} {\bibfnamefont {Soonwon}\ \bibnamefont {Choi}},
  \bibinfo {author} {\bibfnamefont {Vedika}\ \bibnamefont {Khemani}}, \bibinfo
  {author} {\bibfnamefont {Julian}\ \bibnamefont {L{\'e}onard}}, \ and\
  \bibinfo {author} {\bibfnamefont {Markus}\ \bibnamefont {Greiner}},\
  }\bibfield  {title} {\enquote {\bibinfo {title} {Probing entanglement in a
  many-body{\textendash}localized system},}\ }\href {\doibase
  10.1126/science.aau0818} {\bibfield  {journal} {\bibinfo  {journal}
  {Science}\ }\textbf {\bibinfo {volume} {364}},\ \bibinfo {pages} {256--260}
  (\bibinfo {year} {2019})}\BibitemShut {NoStop}%
\bibitem{DaumannARXIV}  Mirko  Daumann and Robin  Steinigeweg and  Thomas  Dahm, ``Many-body localization in translational invariant diamond ladders with flat bands'', (2020) arXiv:2009.09705.
\bibitem [{\citenamefont {Santos}\ \emph
  {et~al.}(2012{\natexlab{b}})\citenamefont {Santos}, \citenamefont
  {Polkovnikov},\ and\ \citenamefont {Rigol}}]{Santos2012PRER}%
  \BibitemOpen
  \bibfield  {author} {\bibinfo {author} {\bibfnamefont {Lea~F.}\ \bibnamefont
  {Santos}}, \bibinfo {author} {\bibfnamefont {Anatoli}\ \bibnamefont
  {Polkovnikov}}, \ and\ \bibinfo {author} {\bibfnamefont {Marcos}\
  \bibnamefont {Rigol}},\ }\bibfield  {title} {\enquote {\bibinfo {title} {Weak
  and strong typicality in quantum systems},}\ }\href@noop {} {\bibfield
  {journal} {\bibinfo  {journal} {Phys. Rev. E}\ }\textbf {\bibinfo {volume}
  {86}},\ \bibinfo {pages} {010102} (\bibinfo {year}
  {2012}{\natexlab{b}})}\BibitemShut {NoStop}%
\bibitem [{\citenamefont {Page}(1993)}]{Page1993}%
  \BibitemOpen
  \bibfield  {author} {\bibinfo {author} {\bibfnamefont {Don~N.}\ \bibnamefont
  {Page}},\ }\bibfield  {title} {\enquote {\bibinfo {title} {Average entropy of
  a subsystem},}\ }\href {\doibase 10.1103/PhysRevLett.71.1291} {\bibfield
  {journal} {\bibinfo  {journal} {Phys. Rev. Lett.}\ }\textbf {\bibinfo
  {volume} {71}},\ \bibinfo {pages} {1291--1294} (\bibinfo {year}
  {1993})}\BibitemShut {NoStop}%
\bibitem [{\citenamefont {Vidmar}\ \emph {et~al.}(2017)\citenamefont {Vidmar},
  \citenamefont {Hackl}, \citenamefont {Bianchi},\ and\ \citenamefont
  {Rigol}}]{Vidmar2017a}%
  \BibitemOpen
  \bibfield  {author} {\bibinfo {author} {\bibfnamefont {Lev}\ \bibnamefont
  {Vidmar}}, \bibinfo {author} {\bibfnamefont {Lucas}\ \bibnamefont {Hackl}},
  \bibinfo {author} {\bibfnamefont {Eugenio}\ \bibnamefont {Bianchi}}, \ and\
  \bibinfo {author} {\bibfnamefont {Marcos}\ \bibnamefont {Rigol}},\ }\bibfield
   {title} {\enquote {\bibinfo {title} {Entanglement entropy of eigenstates of
  quadratic fermionic Hamiltonians},}\ }\href {\doibase
  10.1103/PhysRevLett.119.020601} {\bibfield  {journal} {\bibinfo  {journal}
  {Phys. Rev. Lett.}\ }\textbf {\bibinfo {volume} {119}},\ \bibinfo {pages}
  {020601} (\bibinfo {year} {2017})}\BibitemShut {NoStop}%
\bibitem [{\citenamefont {Alba}\ and\ \citenamefont
  {Calabrese}(2017)}]{Alba2017}%
  \BibitemOpen
  \bibfield  {author} {\bibinfo {author} {\bibfnamefont {Vincenzo}\
  \bibnamefont {Alba}}\ and\ \bibinfo {author} {\bibfnamefont {Pasquale}\
  \bibnamefont {Calabrese}},\ }\bibfield  {title} {\enquote {\bibinfo {title}
  {Entanglement and thermodynamics after a quantum quench in integrable
  systems},}\ }\href {\doibase 10.1073/pnas.1703516114} {\bibfield  {journal}
  {\bibinfo  {journal} {PNAS}\ }\textbf {\bibinfo {volume} {114}},\ \bibinfo
  {pages} {7947--7951} (\bibinfo {year} {2017})}\BibitemShut {NoStop}%
\bibitem [{\citenamefont {Calabrese}(2020)}]{CalabreseARXIV}%
  \BibitemOpen
  \bibfield  {author} {\bibinfo {author} {\bibfnamefont {Pasquale}\
  \bibnamefont {Calabrese}},\ }\href@noop {} {\enquote {\bibinfo {title}
  {Entanglement spreading in non-equilibrium integrable systems},}\ } (\bibinfo
  {year} {2020}),\ \bibinfo {note} {arXiv:2008.11080}\BibitemShut {NoStop}%
\bibitem [{\citenamefont {Wei}\ \emph {et~al.}(2018)\citenamefont {Wei},
  \citenamefont {Ramanathan},\ and\ \citenamefont {Cappellaro}}]{Wei2018}%
  \BibitemOpen
  \bibfield  {author} {\bibinfo {author} {\bibfnamefont {Ken~Xuan}\
  \bibnamefont {Wei}}, \bibinfo {author} {\bibfnamefont {Chandrasekhar}\
  \bibnamefont {Ramanathan}}, \ and\ \bibinfo {author} {\bibfnamefont {Paola}\
  \bibnamefont {Cappellaro}},\ }\bibfield  {title} {\enquote {\bibinfo {title}
  {Exploring localization in nuclear spin chains},}\ }\href {\doibase
  10.1103/PhysRevLett.120.070501} {\bibfield  {journal} {\bibinfo  {journal}
  {Phys. Rev. Lett.}\ }\textbf {\bibinfo {volume} {120}},\ \bibinfo {pages}
  {070501} (\bibinfo {year} {2018})}\BibitemShut {NoStop}%
\bibitem [{\citenamefont {Wei}\ \emph {et~al.}(2019)\citenamefont {Wei},
  \citenamefont {Peng}, \citenamefont {Shtanko}, \citenamefont {Marvian},
  \citenamefont {Lloyd}, \citenamefont {Ramanathan},\ and\ \citenamefont
  {Cappellaro}}]{Wei2019}%
  \BibitemOpen
  \bibfield  {author} {\bibinfo {author} {\bibfnamefont {Ken~Xuan}\
  \bibnamefont {Wei}}, \bibinfo {author} {\bibfnamefont {Pai}\ \bibnamefont
  {Peng}}, \bibinfo {author} {\bibfnamefont {Oles}\ \bibnamefont {Shtanko}},
  \bibinfo {author} {\bibfnamefont {Iman}\ \bibnamefont {Marvian}}, \bibinfo
  {author} {\bibfnamefont {Seth}\ \bibnamefont {Lloyd}}, \bibinfo {author}
  {\bibfnamefont {Chandrasekhar}\ \bibnamefont {Ramanathan}}, \ and\ \bibinfo
  {author} {\bibfnamefont {Paola}\ \bibnamefont {Cappellaro}},\ }\bibfield
  {title} {\enquote {\bibinfo {title} {Emergent prethermalization signatures in
  out-of-time ordered correlations},}\ }\href {\doibase
  10.1103/PhysRevLett.123.090605} {\bibfield  {journal} {\bibinfo  {journal}
  {Phys. Rev. Lett.}\ }\textbf {\bibinfo {volume} {123}},\ \bibinfo {pages}
  {090605} (\bibinfo {year} {2019})}\BibitemShut {NoStop}%
\bibitem [{\citenamefont {Yin}\ \emph {et~al.}(2020)\citenamefont {Yin},
  \citenamefont {Peng}, \citenamefont {Huang}, \citenamefont {Ramanathan},\
  and\ \citenamefont {Cappellaro}}]{Yin2020}%
  \BibitemOpen
  \bibfield  {author} {\bibinfo {author} {\bibfnamefont {Chao}\ \bibnamefont
  {Yin}}, \bibinfo {author} {\bibfnamefont {Pai}\ \bibnamefont {Peng}},
  \bibinfo {author} {\bibfnamefont {Xiaoyang}\ \bibnamefont {Huang}}, \bibinfo
  {author} {\bibfnamefont {Chandrasekhar}\ \bibnamefont {Ramanathan}}, \ and\
  \bibinfo {author} {\bibfnamefont {Paola}\ \bibnamefont {Cappellaro}},\
  }\href@noop {} {\enquote {\bibinfo {title} {Prethermal quasiconserved
  observables in floquet quantum systems},}\ } (\bibinfo {year} {2020}),\
  \bibinfo {note} {arXiv:2005.11150}\BibitemShut {NoStop}%
\bibitem [{\citenamefont {Li}\ \emph {et~al.}(2017)\citenamefont {Li},
  \citenamefont {Fan}, \citenamefont {Wang}, \citenamefont {Ye}, \citenamefont
  {Zeng}, \citenamefont {Zhai}, \citenamefont {Peng},\ and\ \citenamefont
  {Du}}]{Li2017}%
  \BibitemOpen
  \bibfield  {author} {\bibinfo {author} {\bibfnamefont {Jun}\ \bibnamefont
  {Li}}, \bibinfo {author} {\bibfnamefont {Ruihua}\ \bibnamefont {Fan}},
  \bibinfo {author} {\bibfnamefont {Hengyan}\ \bibnamefont {Wang}}, \bibinfo
  {author} {\bibfnamefont {Bingtian}\ \bibnamefont {Ye}}, \bibinfo {author}
  {\bibfnamefont {Bei}\ \bibnamefont {Zeng}}, \bibinfo {author} {\bibfnamefont
  {Hui}\ \bibnamefont {Zhai}}, \bibinfo {author} {\bibfnamefont {Xinhua}\
  \bibnamefont {Peng}}, \ and\ \bibinfo {author} {\bibfnamefont {Jiangfeng}\
  \bibnamefont {Du}},\ }\bibfield  {title} {\enquote {\bibinfo {title}
  {Measuring out-of-time-order correlators on a nuclear magnetic resonance
  quantum simulator},}\ }\href {\doibase 10.1103/PhysRevX.7.031011} {\bibfield
  {journal} {\bibinfo  {journal} {Phys. Rev. X}\ }\textbf {\bibinfo {volume}
  {7}},\ \bibinfo {pages} {031011} (\bibinfo {year} {2017})}\BibitemShut
  {NoStop}%
\bibitem [{\citenamefont {S\'anchez}\ \emph {et~al.}(2020)\citenamefont
  {S\'anchez}, \citenamefont {Chattah}, \citenamefont {Wei}, \citenamefont
  {Buljubasich}, \citenamefont {Cappellaro},\ and\ \citenamefont
  {Pastawski}}]{Sanchez2020}%
  \BibitemOpen
  \bibfield  {author} {\bibinfo {author} {\bibfnamefont {C.~M.}\ \bibnamefont
  {S\'anchez}}, \bibinfo {author} {\bibfnamefont {A.~K.}\ \bibnamefont
  {Chattah}}, \bibinfo {author} {\bibfnamefont {K.~X.}\ \bibnamefont {Wei}},
  \bibinfo {author} {\bibfnamefont {L.}~\bibnamefont {Buljubasich}}, \bibinfo
  {author} {\bibfnamefont {P.}~\bibnamefont {Cappellaro}}, \ and\ \bibinfo
  {author} {\bibfnamefont {H.~M.}\ \bibnamefont {Pastawski}},\ }\bibfield
  {title} {\enquote {\bibinfo {title} {Perturbation independent decay of the
  loschmidt echo in a many-body system},}\ }\href {\doibase
  10.1103/PhysRevLett.124.030601} {\bibfield  {journal} {\bibinfo  {journal}
  {Phys. Rev. Lett.}\ }\textbf {\bibinfo {volume} {124}},\ \bibinfo {pages}
  {030601} (\bibinfo {year} {2020})}\BibitemShut {NoStop}%
\bibitem [{\citenamefont {Ramanathan}\ \emph {et~al.}(2003)\citenamefont
  {Ramanathan}, \citenamefont {Cho}, \citenamefont {Cappellaro}, \citenamefont
  {Boutis},\ and\ \citenamefont {Cory}}]{Ramanathan2003}%
  \BibitemOpen
  \bibfield  {author} {\bibinfo {author} {\bibfnamefont {C.}~\bibnamefont
  {Ramanathan}}, \bibinfo {author} {\bibfnamefont {H.}~\bibnamefont {Cho}},
  \bibinfo {author} {\bibfnamefont {P.}~\bibnamefont {Cappellaro}}, \bibinfo
  {author} {\bibfnamefont {G.S.}\ \bibnamefont {Boutis}}, \ and\ \bibinfo
  {author} {\bibfnamefont {D.G.}\ \bibnamefont {Cory}},\ }\bibfield  {title}
  {\enquote {\bibinfo {title} {Encoding multiple quantum coherences in
  non-commuting bases},}\ }\href {\doibase
  https://doi.org/10.1016/S0009-2614(02)02020-1} {\bibfield  {journal}
  {\bibinfo  {journal} {Chem. Phys. Lett.}\ }\textbf {\bibinfo {volume}
  {369}},\ \bibinfo {pages} {311 -- 317} (\bibinfo {year} {2003})}\BibitemShut
  {NoStop}%
\bibitem [{\citenamefont {Cho}\ \emph {et~al.}(2005)\citenamefont {Cho},
  \citenamefont {Ladd}, \citenamefont {Baugh}, \citenamefont {Cory},\ and\
  \citenamefont {Ramanathan}}]{Cho2005}%
  \BibitemOpen
  \bibfield  {author} {\bibinfo {author} {\bibfnamefont {H.}~\bibnamefont
  {Cho}}, \bibinfo {author} {\bibfnamefont {T.~D.}\ \bibnamefont {Ladd}},
  \bibinfo {author} {\bibfnamefont {J.}~\bibnamefont {Baugh}}, \bibinfo
  {author} {\bibfnamefont {D.~G.}\ \bibnamefont {Cory}}, \ and\ \bibinfo
  {author} {\bibfnamefont {C.}~\bibnamefont {Ramanathan}},\ }\bibfield  {title}
  {\enquote {\bibinfo {title} {Multispin dynamics of the solid-state nmr free
  induction decay},}\ }\href {\doibase 10.1103/PhysRevB.72.054427} {\bibfield
  {journal} {\bibinfo  {journal} {Phys. Rev. B}\ }\textbf {\bibinfo {volume}
  {72}},\ \bibinfo {pages} {054427} (\bibinfo {year} {2005})}\BibitemShut
  {NoStop}%
\bibitem [{Not()}]{NoteCS}%
  \BibitemOpen
  \href@noop {} {}\bibinfo {note} {The central spin initial state is actually $
  \rho^{\text{s}}(0)=\frac{\mathds{1}+\epsilon \sigma_{\text{\tiny X}}}{2}$,
  where $ \epsilon \sim 10^{-5} $ is the nuclear spin polarization at room
  temperature. Since the identity operator does not lead to any observable
  signal, we drop it.}\BibitemShut {Stop}%
\bibitem [{\citenamefont {Haeberlen}(1976)}]{HaeberlenBook}%
  \BibitemOpen
  \bibfield  {author} {\bibinfo {author} {\bibfnamefont {U.}~\bibnamefont
  {Haeberlen}},\ }\href@noop {} {\emph {\bibinfo {title} {High Resolution {NMR}
  in Solids: Selective Averaging}}}\ (\bibinfo  {publisher} {Academic Press},\
  \bibinfo {address} {New York},\ \bibinfo {year} {1976})\BibitemShut {NoStop}%
\bibitem [{\citenamefont {Rhim}\ \emph {et~al.}(1973)\citenamefont {Rhim},
  \citenamefont {Elleman},\ and\ \citenamefont {Vaughan}}]{Rhim73_mrev8}%
  \BibitemOpen
  \bibfield  {author} {\bibinfo {author} {\bibfnamefont {W‐K.}\ \bibnamefont
  {Rhim}}, \bibinfo {author} {\bibfnamefont {D.~D.}\ \bibnamefont {Elleman}}, \
  and\ \bibinfo {author} {\bibfnamefont {R.~W.}\ \bibnamefont {Vaughan}},\
  }\bibfield  {title} {\enquote {\bibinfo {title} {Enhanced resolution for
  solid state nmr},}\ }\href
  {http://aip.scitation.org/doi/abs/10.1063/1.1679423} {\bibfield  {journal}
  {\bibinfo  {journal} {J. Chem. Phys.}\ }\textbf {\bibinfo {volume} {58}},\
  \bibinfo {pages} {1772--1773} (\bibinfo {year} {1973})}\BibitemShut {NoStop}%
\bibitem [{\citenamefont {Gerstein}\ and\ \citenamefont
  {Dybowski}(1985)}]{DybowskiBook}%
  \BibitemOpen
  \bibfield  {author} {\bibinfo {author} {\bibfnamefont {B.~C.}\ \bibnamefont
  {Gerstein}}\ and\ \bibinfo {author} {\bibfnamefont {C.~R.}\ \bibnamefont
  {Dybowski}},\ }\href@noop {} {\emph {\bibinfo {title} {Transient techniques
  in NMR of solids}}}\ (\bibinfo  {publisher} {Academic Press, Inc.},\ \bibinfo
  {year} {1985})\BibitemShut {NoStop}%
\bibitem [{\citenamefont {Somaroo}\ \emph {et~al.}(1998)\citenamefont
  {Somaroo}, \citenamefont {Cory},\ and\ \citenamefont
  {Havel}}]{Somaroo:1998jy}%
  \BibitemOpen
  \bibfield  {author} {\bibinfo {author} {\bibfnamefont {Shyamal~S.}\
  \bibnamefont {Somaroo}}, \bibinfo {author} {\bibfnamefont {David~G.}\
  \bibnamefont {Cory}}, \ and\ \bibinfo {author} {\bibfnamefont {Timothy~F.}\
  \bibnamefont {Havel}},\ }\bibfield  {title} {\enquote {\bibinfo {title}
  {{Expressing the operations of quantum computing in multiparticle geometric
  algebra}},}\ }\href {\doibase 10.1016/S0375-9601(98)00010-3} {\bibfield
  {journal} {\bibinfo  {journal} {Phys. Lett. A}\ }\textbf {\bibinfo {volume}
  {240}},\ \bibinfo {pages} {1--7} (\bibinfo {year} {1998})}\BibitemShut
  {NoStop}%
\bibitem [{\citenamefont {Baum}\ \emph {et~al.}(1985)\citenamefont {Baum},
  \citenamefont {Munowitz}, \citenamefont {Garroway},\ and\ \citenamefont
  {Pines}}]{Baum1985}%
  \BibitemOpen
  \bibfield  {author} {\bibinfo {author} {\bibfnamefont {J.}~\bibnamefont
  {Baum}}, \bibinfo {author} {\bibfnamefont {M.}~\bibnamefont {Munowitz}},
  \bibinfo {author} {\bibfnamefont {A.~N.}\ \bibnamefont {Garroway}}, \ and\
  \bibinfo {author} {\bibfnamefont {A.}~\bibnamefont {Pines}},\ }\bibfield
  {title} {\enquote {\bibinfo {title} {Multiple-quantum dynamics in solid state
  nmr},}\ }\href {\doibase 10.1063/1.449344} {\bibfield  {journal} {\bibinfo
  {journal} {J. Chem. Phys.}\ }\textbf {\bibinfo {volume} {83}},\ \bibinfo
  {pages} {2015--2025} (\bibinfo {year} {1985})}\BibitemShut {NoStop}%
\end{thebibliography}
%


\end{document}